\documentstyle[twoside,fleqn,npb,epsfig]{article}

\newcommand{\AmS}{{\protect\the\textfont2
  A\kern-.1667em\lower.5ex\hbox{M}\kern-.125emS}}
\newcommand{\beq}{\begin{equation}}
\newcommand{\beqa}{\begin{eqnarray}}
\newcommand{\bce}{\begin{center}}
\newcommand{\bfig}{\begin{figure}}
\newcommand{\bit}{\begin{itemize}}
\newcommand{\ben}{\begin{enumerate}}
\newcommand{\eeq}{\end{equation}}
\newcommand{\eeqa}{\end{eqnarray}}
\newcommand{\ece}{\end{center}}
\newcommand{\efig}{\end{figure}}
\newcommand{\eit}{\end{itemize}}
\newcommand{\een}{\end{enumerate}}

\title{Prospects for Higgs properties determination at the 
LHC\thanks{Presented by F.~Piccinini at 
  {\sl RADCOR 2002 and Loops and Legs in Quantum Field Theory}, 
  September 2002, Kloster Banz, Germany.}\thanks{
     The work of FP is supported in part by the EU Fourth Framework Programme
        ``Training and Mobility of Researchers'', Network ``Quantum
        Chromodynamics and the Deep Structure of Elementary
        Particles'', contract FMRX--CT98--0194 (DG 12 -- MIHT). 
ADP is supported by a
      M. Curie fellowship, contract HPMF-CT-2001-01178.}}

\author{F. Piccinini\thanks{On 
leave of absence from INFN Sezione di Pavia, Italy.} 
and
A.D.~Polosa\address{CERN, Theoretical Physics Division, 
                                   CH~1211 Geneva 23, Switzerland}}

\begin{document}

\begin{abstract}
The strategies recently developed to study Higgs boson properties 
at the LHC are reviewed. It is shown how to obtain model-independent 
determinations of couplings to fermions and gauge bosons by 
exploiting different production and decay channels. 
We consider in some detail the case of  Weak Boson Fusion
Higgs production with $H \to b \bar b$ as well as the 
prospects for the determination of the Higgs self-coupling
at the SLHC.
\end{abstract}
\maketitle
\section{Introduction}
The LHC will allow not only the discovery of the Higgs boson, but also the 
study of its properties, such as mass, width and couplings to 
fermions and gauge bosons. While the decay channels $H \to \gamma \gamma$ 
and $H \to Z Z^{(*)} \to 4l$ will allow a direct mass measurement at 
the 0.1\% level over a wide range of masses~\cite{atl99-15}, 
the total width can only 
be determined with about 10\% accuracy by direct measurement of the decay 
$H \to Z Z^{(*)} \to 4l$ for $m_H > 200$~GeV, (the Higgs width 
for lower Higgs masses being too small with respect 
to the detector resolution). 
As it will be shown below, an indirect measurement of the total 
width can be performed also in the low mass region  
by exploiting the available production and decay 
mechanisms at the LHC. 
Several studies have been performed to improve on the strategy 
originally proposed in ref.~\cite{zknrw00} for the determination of the 
Higgs boson properties. Moreover, 
the first analyses of the LHC potential for the Higgs self-coupling 
determination have been worked out. 
We will briefly review the progress recently made in this field.
The main focus will be on the 
mass window 115-200~GeV, which is the one preferred by electroweak 
precision data and partially by supersymmetry. 

\section{Higgs couplings to fermions and gauge bosons}
\label{sec:hcoup}
In principle, the Higgs coupling to a given fermion family $f$, 
could be obtained from the following relation:
\begin{eqnarray}
R(H\to f \bar f)&=&\int{L dt}\cdot \sigma(pp\to H)\cdot 
{\frac{\Gamma_f}{\Gamma}}, \nonumber
\end{eqnarray}
where $R(H\to f \bar f)$ is the Higgs production rate in a given final state, 
$\int{L dt}$ is the integrated 
luminosity, $\sigma(pp\to H)$ is the Higgs production cross section, 
while $\Gamma$ and $\Gamma_f$ are the total and partial Higgs widths 
respectively. A measurement of the Higgs 
production rate in a given channel allows the extraction of the partial
width for that channel which in turn gives the  
coupling $g_{f}$ of the Higgs to
the decay particles involved ($\Gamma_f\sim g_f^2$), provided that the
Higgs production cross-section and the total Higgs width are known
from the theory. 
Aiming at model-independent coupling determinations, one needs to 
consider ratios of couplings, which are experimentally accessible
through the measurements of ratios of rates for different final
states, because the total Higgs cross-section and width
cancel in the ratios 
(as well as the luminosity and all the 
QCD uncertainties related to the initial 
state). 

\bfig[htb]
\bce
\epsfig{file=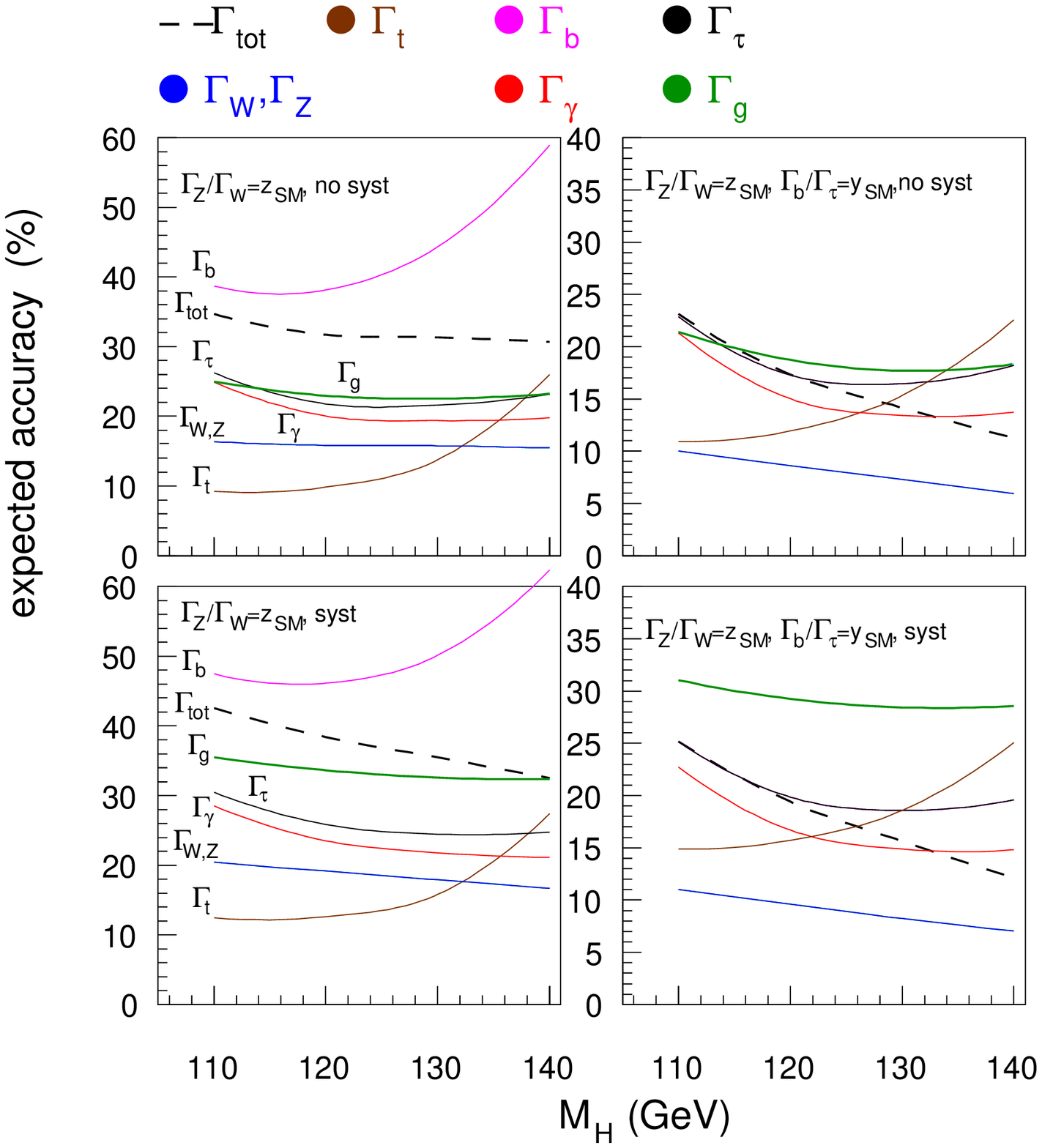,height=7cm,width=7cm}
\caption{\small Relative accuracy (\%) on the individual rates 
$\Gamma_i$ expected at the LHC (from ref.~\cite{rb02}). 
See the text for a detailed description of the panels.}
\label{fig:fig1}
\ece
\efig

In spite of the fact that the gluon fusion mechanism is 
the leading scalar Higgs production mode at the LHC, other subleading 
production modes, such as weak boson fusion and associated production, are 
extremely important to provide complementary information and to allow 
unique determinations of ratios of Higgs boson couplings. 
Up to now detailed studies on signal and backgrounds for several channels 
have been performed, namely 
$gg \to H, (H \to \gamma \gamma, ZZ, 
WW)$~\cite{atl99-15,cms94-38,dreiner,denegri}, 
$qq \to qqH, (H \to \gamma \gamma, \tau \tau, 
WW)$~\cite{rz97,rzh99,prz00,rz99,kprz01}, 
$p p \to t \bar t H, (H \to b \bar b, WW, 
\tau \tau)$~\cite{rws99,dmd01,mrw02,rb02} and 
$p p \to W H, H \to b \bar b$~\cite{dmd02}. Each process depends on two 
Higgs couplings, one from the Higgs boson production and one from the Higgs 
boson decay, with the exception of the weak boson fusion channels, 
for which it is experimentally 
impossible to distinguish between $WW \to H$ and $ZZ \to H$ production 
mechanisms. However, 
since the couplings of a scalar Higgs boson to the $Z$ and $W$ gauge bosons are
closely related by the electroweak $SU(2)$ gauge symmetry, which has been
very successfully tested by the LEP experiments, and since in a large class 
of models the ratio 
of $HWW$ and $HZZ$ couplings is identical to the one in the 
SM, including the MSSM, it is reasonable to rely on the SM value 
$\Gamma_Z / \Gamma_W = z_{SM}$~\footnote{In a very recent paper~\cite{hkyz} 
the importance of the weak boson fusion processes, with $H \to VV$, has 
been pointed out, with the aim of testing possible anomalous $HVV$ couplings.}. 
Under this hypothesis, every production and
decay channel provides a measurement of the ratio $Z_j^{(i)} = 
\Gamma_i \Gamma_j / \Gamma$, where $i = g,W,t$ indicates the
particles involved in the production process while the index 
$j=b,\tau,W,Z,g,\gamma$ is referred to the decay process. In case of 
$m_H < 140$~GeV, the above mentioned channels allow to 
express the individual
rates $\Gamma_t$, $\Gamma_b$, $\Gamma_\tau$, $\Gamma_W$, $\Gamma_g$ and
$\Gamma_\gamma$ as functions of the observables $Z_j^{(i)}$ and of the total 
Higgs width $\Gamma$~\cite{rb02}. 
With the additional assumption that the total 
width is saturated by the known channels $\Gamma = \Gamma_b + \Gamma_\tau 
+ \Gamma_W + \Gamma_Z + \Gamma_g + \Gamma_\gamma$ (otherwise new processes 
would be observed independently of any precision study), an expression for 
$\Gamma$ can be obtained in terms of the measured quantities 
$Z_j^{(i)}$~\cite{rb02}. 
Figure~\ref{fig:fig1}~\cite{rb02} summarizes the relative accuracy on the 
individual rates $\Gamma_i$ expected in the model-independent scenario as 
well as in a scenario with $\Gamma_b/\Gamma_\tau$ fixed to its SM value, 
assuming a total integrated luminosity of 200~fb$^{-1}$. 
The upper plots show the accuracies obtained without including any 
theoretical systematic error, while the lower plots show the same accuracies 
when a systematic theoretical error of $20\%$ for the $gg\to H$ channel, of
$5\%$ for the $qq\to qqH$, and of $10\%$ for the $pp\to t\bar{t}H$
channel are included. As can be seen, the total Higgs width can be indirectly 
determined in the low mass region with a precision of the order of 30\% in a 
model-independent way while 
the Higgs couplings can be determined with 
accuracies between 7\% and 25\%. In the case of $140 < m_H < 200$~GeV, the 
gluon fusion, weak boson fusion and $t \bar t H$ associated production 
processes, with the Higgs 
boson decaying only to gauge bosons, allow an indirect determination of 
$\Gamma_W$ and $\Gamma$ with a precision of the order of 10\%~\cite{zknrw00,z02}. 
In this Higgs mass range, however, there is no handle to study the Higgs Yukawa 
coulings to $b$ quarks and $\tau$ leptons. 
The assumption $\Gamma_Z / \Gamma_W = z_{SM}$ 
can be tested at the 20--30\% level, for $m_H > 130$~GeV, 
by measuring the ratio $Z_Z^{(g)} / Z_W^{(g)}$~\cite{z02}, 
and it can even be tested with the same level of accuracy for lower 
Higgs boson masses by comparing the two ratios $Z_b^{(WH)} / Z_b^{(t)}$ and 
$Z_\tau^{(W)} / Z_\tau^{(t)}$~\cite{rb02}. For $m_H > 140$~GeV, 
with luminosities of the order of 
300~fb$^{-1}$, the ratio $\Gamma_t/\Gamma_g$ can be tested in a 
model-independent way through a measurement of 
$Z_W^{(t)}/Z_W^{(g)}$~\cite{mrw02}. 

\section{$\mathbf{H \to b \bar b}$ via Weak Boson Fusion}
As is clear from Figure~\ref{fig:fig1}, the most poorly known coupling turns 
out to be the $H b \bar b$ Yukawa coupling, 
which can be at best determined with 
an uncertainty at the 20\% level (without any assumption on the ratio 
$\Gamma_b / \Gamma_\tau$). 
To improve the analysis of the $H b \bar b$ Yukawa coupling, one can
consider the decay of an Higgs, produced via Weak Boson 
Fusion, into $b \bar b$ pairs~\cite{hbb}. We report below 
the main results of that study.

Signal and background event estimates are based on a leading order 
partonic calculation of the matrix elements (ME) obtained with the 
event generator ALPGEN~\cite{alpgen}. 
The  background sources considered include:
\begin{enumerate}
\item   QCD production of $b\bar{b}jj$ final states, where $j$ indicates a jet originating from a
light quark ($u,d,s,c$) or a gluon;
\item QCD production of  $jjjj$ final states; 
\item associated production of $Z^*/\gamma^* \to b\bar{b}$ and light
jets, where the invariant mass of the $b\bar{b}$ pair is in the Higgs
signal region either because of imperfect mass resolution, or because
of the high-mass tail of the intermediate vector boson;
\end{enumerate}
along with multiple interaction events ($pp\oplus pp,pp\oplus pp\oplus pp...$) 
giving rise to final states of the kind 
$b\bar{b}jj$ and $jjjj$.
In order to satisfy the requirements of optimization of the signal 
significance, or sensitivity ($S/\sqrt{B}$), and 
compatibility with trigger and data acquisition constraints,
different selection criteria have been considered.
As can be seen  in ref.~\cite{hbb}, 
the 
sensitivity can be as large as $5$ for Higgs 
masses close to the exclusion limit given by LEP searches
but the ratio $S/B$ is only a fraction of a
percent. This implies that the background itself will have to be known
with accuracies at the permille level.  There is no way that this
precision can be obtained from theoretical calculations. The
background should therefore be determined entirely from data. 
The large rate of $b \bar b j j$ from single and multiple interactions 
and the smoothness of their mass distribution in the signal
region will allow to estimate their size with enough statistical accuracy, 
without significant systematic uncertainties. 

The situation is potentially different in
the case of the  backgrounds from the tails of the $Z$
decays. The $Z$ mass peak is sufficiently close to $m_H$, especially in
the case of the lowest masses allowed by current limits, to possibly
distort the $m_{bb}$ spectrum and spoil the ability to accurately
reconstruct  the noise level from data.
These backgrounds rates are at most comparable to the signal at low $m_H$. 
A 10\% determination of these final states, which should be easily achievable
using the $(Z\to \ell^+\ell^-)jj$ control sample and folding in the
detector energy resolution for jets, should therefore be sufficient
to fix these background levels with the required accuracy.

Concerning the multiple interactions, in the simplest case of two 
overlapping events ($pp\oplus pp$), 
there are four possible  combinations of events leading 
to a $b\bar{b}jj$ background:
$(jj)~\oplus~( b \bar{b})$, $(jj)~\oplus~(j_b j_b)$, $(jj_b)~\oplus~(
j j_b)$ and $(b \bar{b})~\oplus~( b \bar{b})$, where $(ab)\equiv pp\to
ab$, and $j_b$ represent a jet given by a light quark or a gluon 
identified as a $b$-jet, because of a mistagging efficiency 
$\epsilon_{fake}$ of the order of 0.01\%--0.05\%. 
A large 
contribution comes from events of the type $(jj_b) \oplus (jj_b)$,
where the $b\bar{b}$ mass spectrum has a broad peak in the middle of
the signal region. The absolute rate of these events (of the order of
the signal rate, when using the lower transverse momentum threshold of
60~GeV) can be determined if the distribution of the beam-line
$z$ vertex separation
between the two overlapping events can be determined with a resolution of
the order of 5-10~mm. These events are significantly reduced in number
when using the higher threshold of 80~GeV for the forward jets.

Table~\ref{tab:ssig60} summarizes the accuracy reachable in the 
${\cal B}(H \to b \bar b)$ and in the $H b \bar b$ Yukawa 
coupling for the case of two different event selections (described 
in detail in ref.~\cite{hbb}), 
assuming that the coupling $HWW$ is the one predicted by the Standard
Model or determined in other reactions studied in the
literature. 
An integrated luminosity of $600$~fb$^{-1}$ 
is considered. 
\begin{table}[h]
\begin{center}
\begin{tabular}{lllll}\hline
& $m_H$~(GeV) & 115  & 120  & 140 \\
\hline
$(a)$ & $\delta \Gamma_b/\Gamma$ & $0.33$  & $0.35$   & $0.71$ \\
 & $\delta y_{Hbb}/y_{Hbb}$ & $0.58$  & $0.51$   & $0.56$ \\
\hline
$(b)$ & $\delta \Gamma_b/\Gamma$ & $0.20$  & $0.19$   & $0.37$ \\ 
 & $\delta y_{Hbb}/y_{Hbb}$ & $0.36$  & $0.30$   & $0.29$ \\ 
\hline
\end{tabular}            
\caption{\label{tab:ssig60}
{\small The statistical significance of the determination of 
the branching ratio $\Gamma_b / \Gamma$ and of the
 $b$-quark Yukawa coupling 
in the configurations (a) and (b) (see ref.~\cite{hbb} for a detailed 
description of the two different event selections), for an 
integrated luminosity of $600$~fb$^{-1}$. 
The $p_{\rm T}^j$ cut on jets 
is $p_{\rm T}^j > 60$~GeV. The case of $p_{\rm T}^j > 80$~GeV, 
presented in ref~\cite{hbb}, doesn't affect sizeably the results. 
Here $\epsilon_{fake}=0.01$.}}
\end{center}
\end{table}
The $H\to b\bar{b}$ decay in the WBF channel could be used 
together with other processes already examined in the literature for a model
independent determination of the ratio of Yukawa couplings
$y_{Hbb}/y_{H\tau\tau}$~\cite{zepp}.

As a conclusion of the analysis presented in ref.~\cite{hbb}, 
the $H\to b\bar{b}$ channel produced in association 
with two jets is suggested as an additional channel to be 
exploited for interesting measurements of the Higgs couplings to fermions.

\section{Higgs self-couplings}
A complete determination of the parameters of the SM would 
require the measurement of the Higgs self-couplings. These include trilinear 
and quadrilinear interactions. In the SM the corresponding 
couplings are fixed at LO in terms of the Higgs mass and vacuum expectation
value $v$, namely $\lambda_{HHH}^{SM} = 3 m_H^2 / v$, 
$\lambda_{HHHH}^{SM} = 3 m_H^2 / v^2$. A direct measurement of $\lambda_{HHH}$ 
could be obtained via the detection of Higgs pair production, where a 
contribution is expected from the production of a single off-shell Higgs which
decays into a pair of Higgses. This contribution is always accompanied by 
diagrams where the two Higgs bosons are radiated independently, with couplings 
proportional to the Yukawa couplings or the gauge couplings. As a result, 
different production mechanisms will lead to different sensitivities of 
the $HH$ rate to the value of $\lambda_{HHH}$. In the literature the following 
SM channels have been considered~\cite{dkmz99}: 
inclusive $HH$ production dominated by the 
partonic process $gg \to HH$; 
vector boson fusion $qq \to qqV^*V^* \to qqHH$, 
associated production with $W$ or $Z$ bosons $q \bar q \to V H H$; 
associated production with top-quark pairs $gg/q\bar q \to t \bar t HH$. 
With the exception of the gluon fusion process, which has a total cross section 
at the level of few tens of fb, the cross section for all other channels is 
of the order of 1~fb over the intermediate Higgs mass range~\cite{dkmz99}. 
Given these low
production rates and the potentially large backgrounds associated to the $HH$ 
final states, a quantitative study of the Higgs self-coupling is very hard 
at the LHC. Recently a study of signal and backgrounds has been performed 
for the $g g \to H H$ channel~\cite{slhc}, both for a standard LHC luminosity 
of $10^{34} \, {\rm cm}^{-2} {\rm s}^{-1}$~\cite{bpr02} and for a possible 
future upgrade of the luminosity to $10^{35}\, {\rm cm}^{-2} 
{\rm s}^{-1}$~\cite{slhc}. Among all possible decay channels, the most 
interesting one turned out to be 
$g g \to HH \to W^+ W^- W^+ W^- \to l^\pm \nu j j l^\pm \nu j j$, 
which has a good branching ratio for $m_H \geq 170$~GeV. The like-sign lepton 
requirement is essential to reduce the high-rate opposite-sign lepton final
states from Drell--Yan and $t \bar t$ production. Potential backgrounds 
to the considered signature are given by $t \bar t + $jets, $W Z + $jets, 
$t \bar t W$, $W W W j j$ including the resonant channel $W(H\to WW)jj$ and 
$t \bar t t \bar t$.
\begin{table*}[htb]
\begin{center}
\begin{tabular}{cccccccc} 
\hline
$m_H $ & Signal & $ t\bar{t} $ & $ W^{\pm} Z $ & 
 $ W^{\pm}W^+W^- $ & $ t\bar{t} W^{\pm} $ & $ t\bar{t} t\bar{t} $ 
& $S/\sqrt{B}$\\
\hline
170~GeV &   350                &  90              & 60             &
   2400              &  1600            & 30   
& 5.4
                          \\ 
200~GeV & 220               & 90             & 60              &
   1500              & 1600           & 30
& 3.8 \\
\hline                          
\end{tabular}
\caption{\small Expected numbers
 of signal and background events after all cuts for the $ gg\to HH\to
  \ 4W \to  l^+ l'^{+} 4j \nu \nu$ final state, for
  $\int {\cal L}=6000$ fb$^{-1}$~\cite{slhc}.}
\label{tab:ggHH}
\end{center}
\end{table*}
By applying the cuts described in ref.~\cite{slhc}, the number of events 
for signal and backgrounds are summarized in Table~\ref{tab:ggHH} for 
an integrated luminosity of 6000~fb$^{-1}$, where a signal significance 
of 5.3 (3.8) $\sigma$ for $m_H = 170 \, (200)$~GeV can be reached, 
optimistically assuming that the main parameters of the detector performance 
will remain the same as those expected at $10^{34}\, {\rm cm}^{-2} 
{\rm s}^{-1}$. 
This would lead to a determination of the total production cross-section 
with a statistical uncertainty of $\pm 20\% \, (\pm 26\%)$ for $m_H = 170$~GeV 
($200$~GeV), allowing a determination of $\lambda_{HHH}$ with statistical 
errors of 19\% (25\%)~\cite{slhc}. In the case of $300$~fb$^{-1}$ only 
the non-vanishing of the Higgs self-coupling could be established 
at 95\% C.L. for $150$~GeV $< m_H < 200$~GeV~\cite{bpr02}. 

\section{Summary}
During the last few years there has been a dramatic improvement in both 
theoretical and experimental studies of several Higgs boson production 
and decay 
channels at the LHC. A strategy has been designed to study, in a
model-independent way, the Higgs couplings to fermions and bosons, which allows 
also, with little theoretical assumption, an indirect determination of the 
total Higgs width. The main results of a very recent analysis of the 
$H\to b \bar b$ channel in Weak Boson Fusion production have been reviewed, 
pointing out its importance for the determination of the $H b \bar b$ Yukawa 
coupling. The potential of the LHC in the determination 
of the Higgs self-coupling has been recently
investigated, but only with an integrated 
luminosity of $6000$~fb$^{-1}$, and in the mass range 
$170 \leq m_H \leq 200$~GeV a quantitative study could be performed. 
\vskip3mm
\noindent
{\bf Acknowledgements}\\
The authors wish to thank F.~Gia\-not\-ti, K.~Jakobs for useful discussions and 
M.L.~Mangano, M.~Moretti and R.~Pittau 
for fruitful collaboration. FP wishes to thank the organizers for 
the kind invitation and for the pleasant atmosphere during the Workshop.

\end{document}